\documentclass[12pt]{article}
\title{\bf The vector coupling in IR region from splittings in bottomonium
}
\author{A.M.Badalian\\
Institute of Theoretical and Experimental Physics,\\
B.Cheremushkinskaya 25, 117218 Moscow,Russia}
\date{}
\begin{document}

\maketitle
\newcommand{\beq}{\begin{eqnarray}}
 \newcommand{\eeq}{\end{eqnarray}}
\newcommand{\be}{\begin{equation}}
 \newcommand{\ee}{\end{equation}}

 \def\la{\mathrel{\mathpalette\fun <}}
\def\ga{\mathrel{\mathpalette\fun >}}
\def\fun#1#2{\lower3.6pt\vbox{\baselineskip0pt\lineskip.9pt
\ialign{$\mathsurround=0pt#1\hfil ##\hfil$\crcr#2\crcr\sim\crcr}}}
\newcommand{\veX}{\mbox{\boldmath${\rm X}$}}
\newcommand{{\SD}}{\rm SD}
\newcommand{\pp}{\prime\prime}
\newcommand{\veY}{\mbox{\boldmath${\rm Y}$}}
\newcommand{\vex}{\mbox{\boldmath${\rm x}$}}
\newcommand{\vey}{\mbox{\boldmath${\rm y}$}}
\newcommand{\ver}{\mbox{\boldmath${\rm r}$}}
\newcommand{\vesig}{\mbox{\boldmath${\rm \sigma}$}}
\newcommand{\vedelta}{\mbox{\boldmath${\rm \delta}$}}
\newcommand{\veP}{\mbox{\boldmath${\rm P}$}}
\newcommand{\vep}{\mbox{\boldmath${\rm p}$}}
\newcommand{\veq}{\mbox{\boldmath${\rm q}$}}
\newcommand{\vez}{\mbox{\boldmath${\rm z}$}}
\newcommand{\veS}{\mbox{\boldmath${\rm S}$}}
\newcommand{\veL}{\mbox{\boldmath${\rm L}$}}
\newcommand{\veR}{\mbox{\boldmath${\rm R}$}}
\newcommand{\ves}{\mbox{\boldmath${\rm s}$}}
\newcommand{\vek}{\mbox{\boldmath${\rm k}$}}
\newcommand{\ven}{\mbox{\boldmath${\rm n}$}}
\newcommand{\veu}{\mbox{\boldmath${\rm u}$}}
\newcommand{\vev}{\mbox{\boldmath${\rm v}$}}
\newcommand{\veh}{\mbox{\boldmath${\rm h}$}}
\newcommand{\verho}{\mbox{\boldmath${\rm \rho}$}}
\newcommand{\vexi}{\mbox{\boldmath${\rm \xi}$}}
\newcommand{\veta}{\mbox{\boldmath${\rm \eta}$}}
\newcommand{\veB}{\mbox{\boldmath${\rm B}$}}
\newcommand{\veH}{\mbox{\boldmath${\rm H}$}}
\newcommand{\veE}{\mbox{\boldmath${\rm E}$}}
\newcommand{\veJ}{\mbox{\boldmath${\rm J}$}}
\newcommand{\veal}{\mbox{\boldmath${\rm \alpha}$}}
\newcommand{\vegam}{\mbox{\boldmath${\rm \gamma}$}}
\newcommand{\vepar}{\mbox{\boldmath${\rm \partial}$}}
\newcommand{\llan}{\langle\langle}
\newcommand{\rran}{\rangle\rangle}
\newcommand{\lan}{\langle}
\newcommand{\ran}{\rangle}

\begin{abstract}
The splittings $1D-1P, 1F-1P, 2D-2P$ are shown to be the most
convenient characteristics to determine the vector coupling in IR
region.  In background perturbation theory the splitting $\Delta
=1D -1P$ appears to be in agreement with $\Delta (\exp) =261.1 \pm
2.2$ MeV only for large freezing value, $\alpha_{crit} \cong
0.60$, which corresponds to $\Lambda_{\overline{MS}} (2$-loop,
$n_f=5)\cong 240$ MeV $(\alpha_s(M_Z) =0.1193(2)$. The masses
$M_{cog} (1F)=10362\pm 3$ MeV and $M_{cog} (2D) =10452\pm 4 $ MeV
are predicted.

\end{abstract}

1. The static potential plays a special role in heavy quarkonia
physics. Although $V_{st}(r)$ was introduced by the Cornell group
30 years ago \cite{1}, even now we do not fully understand some
important features of static interaction in QCD, moreover
uncertainties refer both to perturbative (P) and nonperturbative
(NP) contributions to $V_{st}(r)$.  There are three characteristic
features of  static potentials, widely  used in QCD
phenomenology:\\
 1.{\underline {additivity}}, when $V_{st}(r)$ is taken as a sum of
confining  and the gluon-exchange terms:
\be
V_{st}(r) = V_{NP} (r) + V_{GE} (r);\label{1}\ee
2.{\underline{linear}} behavior of $V_{NP} (r) =\sigma r$ over the
whole  region of $Q\bar Q$ separations $r$;
3.{\underline{constant}} value of the vector coupling
$\alpha_V(r)$ at large  $r$ \cite{2,3}.
(By definition $V_{GE} (r) =-\frac43 \frac{\alpha_V(r)}{r}$). In
some cases $\alpha_V(r)=$const. is taken  already at not large
$r~~(r\ga 0.2$ fm), as in lattice QCD \cite{4},  or even  at any
separations $r$ as in the Cornell potential \cite{5}. Last
assumption can be justified only if the "true" vector coupling
freezes at rather small  $r$, while asymptotic freedom  behavior
is supposed to be inessential in first approximation, as  for high
excitations in charmonium.

Although for long  time  "the freezing" of $\alpha_V(r)$ is widely
used in QCD phenomenology,  nevertheless, till now there is no
consensus about the true value of freezing (or critical) constant
$\alpha_{cr}$. In different  approaches the values of
$\alpha_{cr}$ vary from $\alpha_{cr} (lat) \approx 0.23\div 0.30$
in lattice QCD \cite{4}, the values 0.39-0.45 for the Cornell
potential \cite{5} up to the number  $\alpha_{crit} \approx 0.60$
in background perturbation theory (BPT)  \cite{3,6} and also in
the famous paper \cite{2}. Even larger values, $\alpha_{eff} (1
GeV) \cong 0.9 \pm 0.1$, were determined from the hadronic decays
of the $\tau$-lepton \cite{7} and in analytical perturbation
theory (APT) \cite{8}, where $\alpha_{cr}(n_f=3)\cong 1.4$.

Some achievements in our understanding  of static interaction on
the fundamental level  mostly refer to NP term. In particular,
{\underline the property of additivity} has been confirmed by
lattice calculations of static potentials  in different group
representations where the behavior $V_{st} (r, N) = C_F
\tilde{V}_{st}$ (universal) $(r\la 1$ fm), or the Casimir
\underline{scaling} property, has been observed in \cite{9}, and
in Ref. \cite{10} theoretical interpretation of the Casimir
scaling  has been  given.

With the use of   the vacuum correlators, measured on the lattice
\cite{11}, it was shown that linear behavior of $V_{NP}(r)$ takes
place only in the range $T_g\la r \la R_{SB} $ \cite{12}. Here
$T_g\approx 0.2$ fm is the gluonic correlation length \cite{11},
while $R_{SB} \approx 1.2$ fm characterizes those separations
$r>R_{SB}$, where the string breaking is  essential \cite{13,14}
and linear
 potential is becoming more flat.
Such flattening of confining potential does not affect the
positions of the $b\bar b$ levels, which lie below $B\bar B$
threshold, but this effect is essential for the states of large
size : $\sqrt{\lan r^2\ran}_{nL}\ga 1.2$ fm. In particular, in
light meson  sector this  effect provides a correlated large shift
(down) of radial excitations  like $\rho(3S), \rho(4S), a_J(2P)$
\cite{14}.

At present there is no     theory of  string breaking and  also we
do not know  precise behavior of $V_{NP}(r)$ at small $r$.
Meanwhile, knowledge of $V_{NP}(r)$ at small $r$  is very
important: (\ref{1}) for understanding   of fine structure
splittings of $\chi_b$ mesons (through the Thomas precession term)
\cite{15}; (\ref{2}) for explanation of very small  shift of
$h_c(1^1P_1)$ with respect to $M_{cog}(1^3P_J)$ in charmonium,
where a cancellation of two small terms--negative $P$ term and
positive NP term takes place \cite{16}.

Here we concentrate on the gluon-exchancge term. A unique
information about the vector coupling $\alpha_V(r)$ can be
extracted from the analysis of the splittings between high
excitations (still lying below $B\bar B$ threshold) in
bottomonium. There are several reasons for that. First, the
splittings between $b \bar b$ levels are known from experiment
with precision accuracy,  $\delta M\la 1$ MeV. Second, there are
\underline{ten} observed (plus unobserved $1F, 2D, 3D,$ and may be
$ 3P)$ states which lie below $B\bar B$ threshold. These states
have very different r.m.s. radii, which spread from 0.2 fm for
$\Upsilon(1S)$ up to 0.8 fm for $2D$ and $3P$ states (see Table
1).

\vspace{0.5cm}

\begin{tabular}{lp{11cm}}
{\bf Table 1}:& The r.m.s. radii $\sqrt{\lan r^2\ran_{nL}}$ in
bottomonium from \cite{6}
\end{tabular}

\begin{center}
\begin{tabular}{|c|c|c|c|c|c|c|c|c|c|}\hline
&&&&&&&&&\\
state& 1S& 1P& 2S& 2P& 1D& 1F& 3S& 2D& 3P
\\ &&&&&&&& &\\\hline &&&&&&&& &\\
$\sqrt{\lan
r^2\ran_{nL}}$&0.22&0.38&0.46&0.62&0.62&0.63&0.71&0.73&0.82
\\
in fm&&&&&&&&&\\\hline
\end{tabular}
\end{center}

In our analysis of  the $b\bar b$ spectrum we use $V_B(r)=\sigma r
-\frac43 \frac{\alpha_B(r)}{r}$ ~~where  the vector coupling
$\alpha_B (r) $  is defined as in BPT \cite{3,6},
\be
\alpha_B(r)=\frac{2}{\pi} \int^\infty_0 \frac{dq}{q} \sin (qr)
\alpha_B (q),\label{2} \ee while the background coupling in
momentum space is given by the standard formula:
\be
\alpha_B(q) =\frac{4\pi}{\beta_0 t_B}
\left(1-\frac{\beta_1}{\beta_0^2} \frac{\ln
t_B}{t_B}\right)\label{3}\ee with  the modification of the
logarithm: \be t_B(q) =\ln
\frac{q^2+M^2_B}{\Lambda^2_V},\label{4}\ee where $ M_B=2.24(1)
\sqrt{\sigma}$ is so called  background mass, defined by  the
lowest hybrid excitation and expressed through  the string tension
\cite{3}. The QCD constant $\Lambda_V$ is expressed through the
conventional $\Lambda_{ \overline{MS}}$ \cite{18}:
\be
\Lambda_V(n_f) =\Lambda_{\overline{MS}}(n_f) \exp
\frac{a_1}{2\beta_0},~~ a_1=\frac{31}{3} -\frac{10}{9}
n_f.\label{5}\ee

The important feature of the background coupling $\alpha_B(q)$
(and also $\alpha_B(r)$) is that it has correct perturbative limit
at large $q^2$ (small $r$). Therefore in BPT \underline{there are
no  additional (fitting) parameters} and the $b\bar b$ spectrum
and the wave functions are fully defined by the QCD constant
$\Lambda_{\overline{MS}} (n_f)$ and the string tension (in BPT the
pole mass of a heavy quark coincides with the conventional value
\cite{18}).

Due to the $r$-dependence of $\alpha_B(r)$ every $b\bar b$ state
has its own characteristic coupling, denoted  as
$\alpha_{eff}(nL),$ which can be defined as \be \alpha_{eff}(nL)
\lan r^{-1}\ran_{nL} = \lan \alpha_B(r)
r^{-1}\ran_{nL}.\label{6}\ee Their values are smaller by $(20\div
30)\%$ than the freezing  value $\alpha_{cr} =\alpha_B (q^2=0)$
and grow for  higher excitations  (see Table 2).

\vspace{0.5cm}

\begin{tabular}{lp{11cm}}
{\bf Table 2}:& The effective coupling $\alpha_{eff} (nL) $ from
\cite{6}
\end{tabular}

\begin{center}
\begin{tabular}{|c|c|c|c|c|c|c|c|c|}\hline
&&&&&&&&\\
state& 1S& 2S& 3S& 1P&  2P& 1D& 2D& $\alpha_{cr}$\\ &&&&&&&&
\\\hline &&&&&&&& \\
$\alpha_{eff}(nL)$&0.41&0.45&0.46&0.50&0.51&0.54&0.54&0.60
\\
&&&&&&&&\\\hline
\end{tabular}
\end{center}

The picture is different for the Cornell potential where
$\alpha_V(r) =const\approx 0.4 $ for all $b\bar b$ states
\cite{5}.

It is convenient to consider \underline{the splittings} between
the spin-averaged masses $M_{cog}(nL)$ (instead of the  absolute
masses), in this way minimizing dependence of the splittings on
the choice of parameters present in  $V_{st}(r)$. It appears that
the splittings $1D-1P, 1F-1P, 2D-2P$ do not practically depend on
kinematics, see Table 3, where the solutions for relativistic
Spinless Salpeter Equation (SSE) are compared  to those for
Schroedinger eq. From Table 3 one can also see that other
splittings (like 2S-1P, 2P-1P) in NR case turn out to be by 6-10
MeV larger and this difference is much larger than the
experimental error in a splitting, $\delta M\sim 1$ MeV.
\newpage

\vspace{0.5cm}

\begin{tabular}{lp{11cm}}
{\bf Table 3}:& The splittings $\Delta=M_{cog} (n_1L_1) - M_{cog}
(n_2 L_2) $ (in MeV) in nonrelativistic (NR) case and for SSE (R
case) $^a)$
\end{tabular}

\begin{center}
\begin{tabular}{|c|c|c|c|c}\hline
&&&&\\
splitting& NR& R&experiment&
\\ &&&&\\\hline &&&&\\
2S-1P& 135&125&$\la 123$\\ 2P-1P&377&370&360$\pm $1.2\\
3S-1P&111&103&$\la 95$ \\ 2D-1P&569&562&-&\\ &&&&\\\hline &&&&\\
1D-1P& 259&259&261.1$\pm$ 2.2\\ 2D-2P&192 &192& -&\\
1F-1P&462&462&-&\\\hline

\end{tabular}
\end{center}
$^a)$ The potential $V_B(r)$ in BPT is taken with $\sigma  =
0.178$ GeV$^2$, $M_B=0.95$ GeV, $\Lambda_V(2$-loop, $n_f=5)=330$
MeV, or $\Lambda_{\overline{MS}}(2$-loop, $n_f=5)=242$ MeV.
\vspace{1cm}

Our calculations  show that the splittings $1D-1P, 1F-1P$, and
$2D-2P$  do not  practically depend on the variation of the quark
pole mass,  kinematics and weakly depend on the variation of the
string tension. In \cite{6} it has been shown that only the value
of  $\sigma =0.177 (3) $ GeV$^2$ provides good description of
bottomonium spectrum as a whole. However these splittings appear
to be very sensitive to the  freezing  value $\alpha_{cr}$ or to
the QCD constant $\Lambda$.

This statement is illustrated by the numbers presented in Table 4
for three values of $\Lambda_V(n_f=5)= 300$ MeV, 320 MeV, and 330
MeV which correspond to two-loop
$\Lambda_{\overline{MS}}(n_f=5)=220$ MeV, 234 MeV, and 242 MeV,
respectively.

\newpage

\vspace{0.5cm}

\begin{tabular}{lp{11cm}}
{\bf Table 4}:& The splittings  between spin-averaged masses in
bottomonium (in MeV) for Spinless Salpeter Equation ($R$ case) for
the  potential  $V_B(r)$ with $\sigma =0.178$ GeV$^2$, $m_b(pole)
\cong 4.83$ GeV, $M_B=0.95$ GeV.
 \end{tabular}

\begin{center}
\begin{tabular}{|c|c|c|c|}\hline
&&&\\
splitting& $\Lambda_V^{(5)} =300$ MeV,& $\Lambda_V^{(5)}=320$
MeV,& $\Lambda_V^{(5)}=330$ MeV,
\\ & $\Lambda^{(5)}_{\overline{MS}}=220$ MeV&
 $\Lambda^{(5)}_{\overline{MS}}=234$ MeV & $\Lambda^{(5)}_{\overline{MS}}=242    $ MeV\\
&&&\\\hline &&&\\1D-1P& 252&257& 259\\ 1F-1P& 450&457& 461\\
2D-2P&188&190&192\\ 2S-1P&123&124&125\\ 3S-2P&103&103&103\\ \hline

\end{tabular}
\end{center}

From Table 4  it is clear  that the splitting $\Delta =M_{cog}
(1D)-M_{cog}(1P)$ turns out to be in good agreement with the
experimental number, $\Delta (\exp) =261.1\pm 2.2
(\exp)^{+1}_{-0}(th)$ MeV only for large value of the QCD constant
$\Lambda^{(5)}_{\overline{MS}}(2$-loop) (experimental number for
$M(1^3 D_2) =10161.1 \pm 2.2.$ MeV is taken from \cite{19}). For
$\Lambda^{(5)}_V(2$-loop)$\approx 335$ MeV the critical value of
$\alpha_B(r)$ is large, $\alpha_{cr}=0.60\pm 0.01$ and
corresponding $\Lambda^{(5)}_{\overline{MS}}(2$-loop)$\approx
240\div 245$ MeV gives rise to $\alpha_s(M_Z)=0.1193(2)$.

From this analysis we can predict $1F-1P,2D-2P$ splittings (or the
masses of $1F$ and $2D$ states) taking  the same $\Lambda_V^{(5)}$
as for the $1D$ state. It gives
\be
M_{cog}(1F) =10362\pm 2 (\alpha_V)\pm 1(\sigma) {\rm
MeV}\label{6}\ee $$ M_{cog}(2D) =10452\pm 2 (\alpha_V)\pm
2(\sigma) {\rm MeV}$$

Since fine structure splittings of the $1^3F_J, 2^3D_J$ multiplets
should be very small, as well as for $1^3D_J$ states \cite{20},
one can expect that the  masses $M(1^3F_J)$ and $M(2^3D_J)$ have
to be very close to the  figures given in (\ref{6}). Therefore the
observation of the $1F, 2D$ states would be crucially important
for the  better understanding of the gluon-exchange term on the
fundamental level.

\end{document}